# Modelling user behavior towards smartphones and wearable technologies: A bibliometric study and brief literature review

1st Maral Jamalova [1]*
[1] Azerbaijan State University of Economics (UNEC), Baku, Azerbaijan
Maral_jamalova@unec.edu.az

**Abstract**—The study uses bibliometric as well as content analysis to determine the current situation regarding the application of technology adoption models (i.e., the Technology Acceptance Model, Unified Theory of Acceptance and Use of Technology, and Innovation Diffusion Theory) to the smartphone market that also includes smart wearables. Hereby the author would like to determine the connection between smartphone usage and adoption models and enrich literature by defining state-of-the-art tendencies and approaches. To achieve the goal, the author applied a two-stage approach: in the first stage, 213 articles were analyzed using Citation and Bibliographic coupling tools in VOSviewer (1.6.20). The papers were selected from the Scopus database and the search of the papers was conducted in the fields of Economics, Business, and Computer technologies. In the second stage, the author conducted a brief literature review of the most influential papers. The results illustrate the situation regarding the implementation of different models in the case of smartphone adoption. Content analyses of the most influential papers were applied to explain and enrich the results of bibliometric analyses as well as determine research gaps and future research development.

**Keywords—** technology acceptance models, smartphone, bibliometric analysis, TAM, UTAUT/UTAUT2, IDT.

## 1 Introduction

Smartphones became one of the essential parts of our life after 2007. Numbers regarding smartphone diffusion increased significantly, as well as the areas where these gadgets could be applied have changed over the last 15 years. Nowadays, everyone uses smartphones for NFC payments, purchasing tickets as well as participating in meetings, making notes, and much more. All these different applications of handsets changed perception towards it; moreover, with the rapid development of technologies, the smartphone became a crucial point as customers began to connect it with smart wearable devices such as smart watches/ fitness bands, headsets, or earbuds. The mentioned changes in technology increased the necessity towards understanding the purchase decision of the products. As a result, the author of the current study attempts to understand and combine the best indicators to define the best model for measuring attitudes toward smartphones.

The purpose of the study is to present an overall summary of the current and previous studies regarding the application of three technology adoption models (Technology Acceptance Model, Unified Theory of Acceptance and Use of Technology, and Innovation Diffusion Theory) on smartphone usage and diffusion. The author is interested in determining the usage of the three most famous technology adoption models in the context of smartphones as well as determining the most influential papers regarding each case. Using the knowledge obtained from bibliometric analyses, the author will carry out a brief literature review based on the highly cited publications. The author looks for answers regarding the following research questions in the case of each model:
1. How should the most influential papers have been mapped using VOSviewer?
2. How have the papers been connected via Citation linkage?
3. How can bibliographic coupling be applied to visualize the relationships between journals?
4. How the development direction of the studies can be generalized?

The result of the two-stage analyses is to provide a profound understanding of modern discussions and development in the field of research concerning practical and theoretical applications of the Technology Acceptance Model (i.e., TAM), Unified Theory of Acceptance and Use of Technology (i.e., UTAUT) and Innovation Diffusion Theory (i.e., IDT) in respect to smartphone diffusion and usage.

## 2 Literature review

A review of the literature illustrates that TAM, UTAUT/UTAUT2, and IDT are three main models mostly applied in the case of smartphones and portable/wearable devices [1], [2]. All of the above-illustrated models are the models that focus on voluntary acceptance of technology and are classified as technology acceptance models.

### 2.1 Technology Acceptance Model - TAM

TAM was introduced by Davis in 1986 and is known as one of the frequently used models in the field of IS. The model was based on the two influential models originating from psychology which are the Theory of Reasonable Action (TRA) and the Theory of Planned Behavior (TPB). TRA was proposed by Ajzen and Fishbein [3] attempting to explain rational behavior which is based on the utilization of information. The theory mostly focused on explaining behavioral intention rather than attitudes. It had several limitations [4] that Ajzen wanted to address by involving perceived behavioral control. As a result, he developed a new model – TPB [5] for explaining behavior even though a person has "incomplete volitional control" [6].

Early versions of the model consider perceived usefulness, perceived ease of use, and attitudes as the main determinants of behavior explaining voluntary use of any kind of technology [7]. In this version perceived usefulness and perceived ease of use had a direct impact on the formulation of attitude [8]. However, later the author provided evidence of an incomplete relationship that resulted in the removal of attitude from the model [9]. Afterward, Davis and his colleagues included some new variables and extended TAM [10], [11].

### 2.2 Unified Theory of Acceptance and Use of Technology (UTAUT/UTAUT2)

The early version of the UTAUT was derived from eight theories and was designed to determine the main variables affecting employees' adoption of IT. However, Venkatesh et al [12] were the first authors who took into account the impact of moderators (i.e., age, gender, experience, and voluntariness of use) in the context of technological products. The original version of the model included independent variables such as Performance and Effort expectancy, Social Influence, Facilitating Conditions as well as above mentioned moderators [12].

The extension of the model (i.e., UTAUT2) was proposed in 2012 [13] and differs from the original version by including additional variables like Hedonistic Motivation, Price per Value, and Habit as well as moderatos [14]. The model targets to identify technology usage from the customer/end-user perspective. The customer/end-user perspective as well as the involvement of moderators are two main points that make UTAUT2 one of the competitors of TAM. Some researchers claim that even though the models include moderators, these variables are not commonly used in the research process [14], [15].

### 2.3 Innovation Diffusion Theory – IDT

Innovation Diffusion Theory (IDT) was introduced by Rogers in 1962 and is known as one of the most frequently used models in the field of Information Systems [16]. The model was based on the relationship between the five variables and attitude toward technology use.

These variables are relative advantage, compatibility, complexity, trialability, and visibility which are combined under the perceived attributes of innovations definition. Unlike earlier explained models, Innovation Diffusion Theory [16] was developed as a component of innovation management to assess the utilization of new technologies. The main weakness that might impact the usage of IDT is connected with the attitude-related weakness of the model [17]. Also, some authors illustrate that the model lacks a connection between innovation properties and expected attitudes [18]. The application of moderators might increase the exploratory power of the model [15] and increase the body of knowledge with new information.

## 3 Research methodology: the two-stage approach.

The application of bibliometric analysis (i.e., performance analysis) to Information Technology and Information System research has rapidly grown and focused on the impact of research quality, author influence, qualification, and impact of the journal and organization in the selected field. It has also been used extensively to understand the Internet of Things-related (including smartphone) situations on specific fields or topics. In the first stage, the author chose the studies published in the Scopus database between 2010 and 2022 that were focused on the application of different models related to smartphone diffusion/adoption. The main purpose of choosing the Scopus database is related to the largest number of publications [19] in comparison to other databases (i.e., WOS). To be able to generalize previous findings the author uses only two types of performance analyses [20] which are named citation analyses (including citation linkage) and bibliographic coupling.

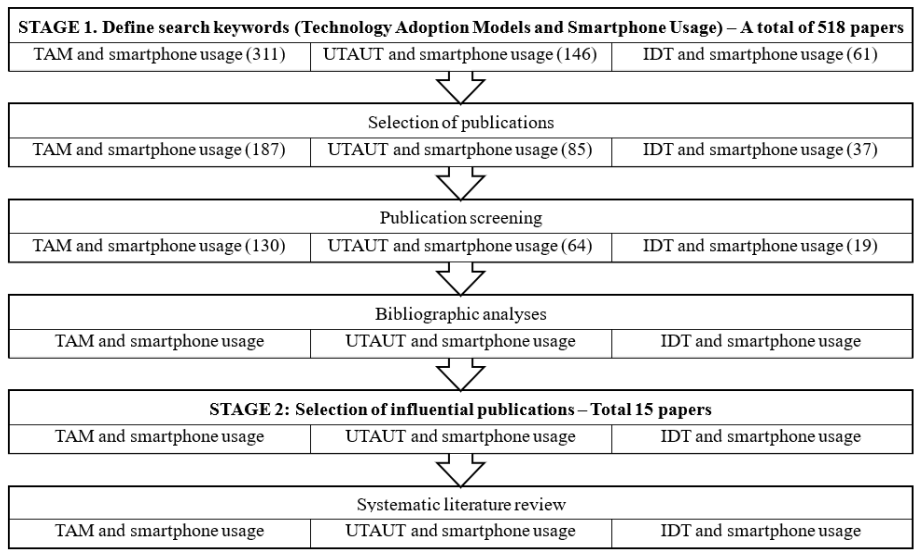

**Fig. 1.** Structure of analyses

Source: own editing

The first stage was focused on defining search keywords, selecting and screening publications, as well as conducting bibliographic analyses. However, the second stage concentrated on the selection of influential publications and a brief literature review.

The below-illustrated search line was used to obtain the information regarding the title, keywords, and abstracts of all publications in the Scopus collection. After analyzing the results, the author decided to keep only articles and limit publications to the English language. So, only 187 publications remained. After screening 9 papers were deleted as they do not directly relate to the discussed topic. Some of the publications (i.e., 48 papers) used more than one model, hereby they were excluded from analyses.

In the case of the Technology Acceptance Model (i.e., TAM):

*SUBJAREA ( busi ) OR SUBJAREA ( econ ) OR SUBJAREA ( comp ) AND TITLE-ABS-KEY ( ( "TECHNOLOGY ACCEPTANCE MODEL" OR tam OR tam2 OR tam3 ) AND ( smartphone OR "smart phone" ) ) AND ( EXCLUDE ( PUBYEAR , 2023 ) OR EXCLUDE ( PUBYEAR , 2009 ) OR EXCLUDE ( PUBYEAR , 2008 ) OR EXCLUDE ( PUBYEAR , 2007 ) )*

Only 85 out of 146 papers were used for analyzing UTAUT/UTAUT2 for the smartphone and/or wearable device market. 64 articles remained after screening and the exclusion of the papers that used more than one model. In the case of the Unified Theory of Acceptance and Use of Technology (i.e., UTAUT):

*SUBJAREA ( busi ) OR SUBJAREA ( econ ) OR SUBJAREA ( comp ) AND TITLE-ABS-KEY ( ( utaut OR utaut2 OR "unified theory of acceptance and use of technology" OR "Extension of Unified Theory of Acceptance and Use of Technology" ) AND ( smartphone OR "SMART PHONE" ) ) AND ( EXCLUDE ( PUBYEAR , 2023 ) OR EXCLUDE ( PUBYEAR , 2009 ) OR EXCLUDE ( PUBYEAR , 2008 ) OR EXCLUDE ( PUBYEAR , 2007 ) )*

Only 37 out of 61 papers were used for analyzing IDT for the smartphone and/or wearable device market 19 articles remained after screening and the exclusion of the papers that used more than one model. In the case of Innovation Diffusion Theory (i.e., IDT):

*SUBJAREA ( busi ) OR SUBJAREA ( econ ) OR SUBJAREA ( comp ) AND TITLE-ABS-KEY ( ( "innovation diffusion theory" OR idt OR "diffusion of innovations" ) AND ( smartphone OR "smart phone" ) ) AND ( EXCLUDE ( PUBYEAR , 2023 ) OR EXCLUDE ( PUBYEAR , 2009 ) OR EXCLUDE ( PUBYEAR , 2008 ) OR EXCLUDE ( PUBYEAR , 2007 ) OR EXCLUDE ( PUBYEAR , 2006 ) )*

## 4 Descriptive analysis

The figure illustrated below (Figure 2) shows the number of publications that were published each year using one of the three most popular technology adoption models. It is clear that year by year the smartphone became a more popular tool and the number of publications in the field began also to increase. Even if TAM was in a leading position from the beginning, the number of UTAUT-based publications increased significantly in a Scopus database after 2018.

Table 1 illustrates the journals that published the highest number of scientific papers about smartphones and wearables. In the mentioned table, SJR is an abbreviation of SCImago Journal Rank and illustrates the positions of different journals based on weighted citations. The higher number of SJRs is also closely related to the popularity of the journal. The Journal of Theoretical and Applied Information Technology published the highest number of publications related to smartphone/ wearable device adoption while SJR for 2022 of the serials is not very high. Computers in Human Behavior, Journal of Retailing and Consumer Services, and Cyberpsychology,

Behavior, and Social Networking are the journals with the SJR scores, and each of the journals published seven, four, and five papers respectively.

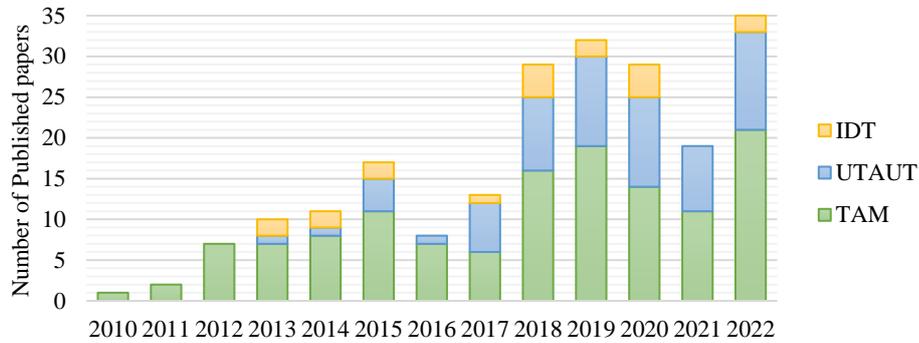

**Fig. 2.** Papers published on smartphone adoption using TAM/UTAUT/IDT (2010–2022); Source: own editing.

**Table 1.** The journals that published the highest number of scientific papers about smartphones and wearables

| N | Journals | SJR 2022 | TAM | UTAUT / UTAUT2 | IDT | Total |
|---|---|---|---|---|---|---|
| 1 | Journal of Theoretical and Applied Information Technology | 0.165 | 5 | 4 | 0 | 9 |
| 2 | Computers in Human Behavior | 2.464 | 3 | 3 | 1 | 7 |
| 3 | International Journal of Mobile Communications* | 0.555 (2019) | 0 | 5 | 0 | 5 |
| 4 | Cyberpsychology, Behavior, and Social Networking | 1.466 | 4 | 0 | 0 | 4 |
| 5 | Journal of Retailing and Consumer Services | 2.543 | 3 | 0 | 1 | 4 |
| 6 | International Journal of Interactive Mobile Technologies | 0.409 | 3 | 0 | 1 | 4 |
| 7 | International Journal of Technology and Human Interaction | 0.189 | 3 | 0 | 0 | 3 |
| 8 | Cogent Business and Management | 0.524 | 3 | 0 | 0 | 3 |
| 9 | International Journal of Recent Technology and Engineering* | 0.107 (2019) | 2 | 1 | 0 | 3 |
| 10 | Journal of Advanced Research in Dynamical and Control Systems * | 0.129 (2019) | 2 | 1 | 0 | 3 |

Note: * -coverage discontinued in Scopus; Source: own editing

## 5 Results – Stage 1: Bibliographic analyses of the selected studies

### 5.1 Citation of documents - TAM; UTAUT/UTAUT2; DOI

The citation is a tool that allows the reader to determine how much a scientific paper, author, or journal impacts the creation of the body of knowledge [20]. In bibliometric research, different units of analysis might be chosen depending on the purpose of the researcher. The documents were selected as the unit of analysis in the current study to find an answer to the first research question. In the case of all three models, the author selected two as a minimum number of citations per scientific work for this analysis. As a result, 102 out of 130 TAM-related papers were in line with the threshold and were combined in 75 clusters (Figure 3). Moreover, the clusters below are not connected. Once again it illustrates that the research in the mentioned field is conducted randomly. Based on Figure 3, the most cited papers were published between 2010 and 2017.

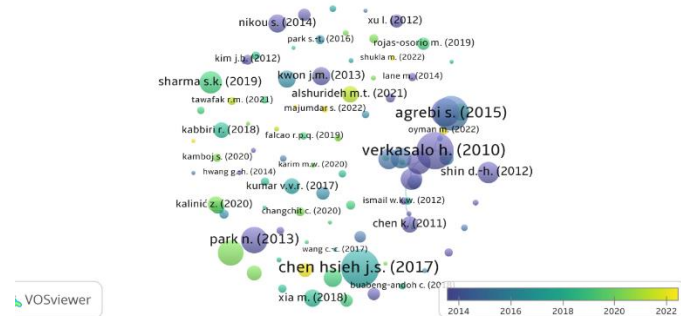

**Fig. 3.** Citation of TAM–related scientific works: visualization of 102 studies

In the case of UTAUT/UTAUT2, 69 out of 85 papers were in line with the threshold (i.e., number of citations = 2) and were combined in 50 clusters (Figure 4). Moreover, the clusters below are not connected.

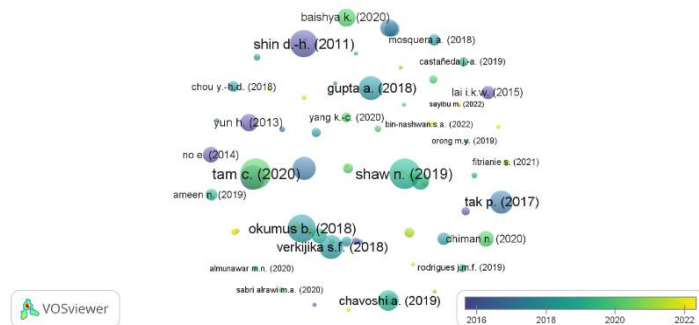

**Fig. 4.** Citation of UTAUT/UTAUT2–related scientific works: visualization of 69 studies

In the case of IDT, 15 out of 19 papers were in line with the threshold (i.e., number of citations = 2) and were combined in 14 clusters (See Figure 5). Mostly, the clusters below are not connected. Nevertheless, the paper of Pham and Ho from 2015 is one of the most cited (i.e., 171 citations) and influential works concerning NFC payments. The other prominent paper which was published by Kaur et. al. [21] also adapted IDT to the mobile payments/ wallets' topic (i.e., 70 citations).

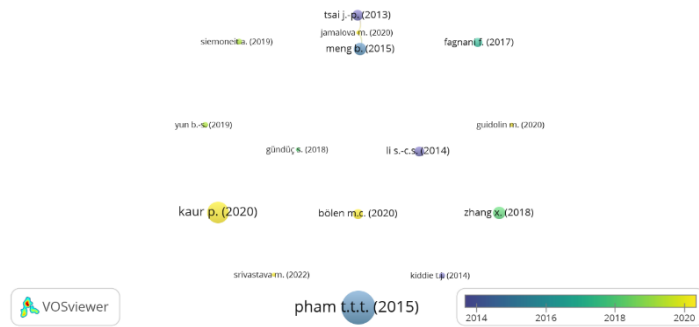

**Fig. 5.** Citation of IDT–related scientific works: visualization of 15 studies

### 5.2 Citation linkage of documents - TAM; UTAUT/UTAUT2; DOI

The citation linkage of documents determines how the scientific papers are interconnected with each other. This section tries to find an answer to the second research question. In the case of TAM, the largest set of documents consists of 23 scientific works that might be combined into seven categories; it is about 22.5% of the overall sample (Figure 6). The authors of these studies are illustrated in the figure below. Interestingly, this group included only the study of Agrebi and Jallais from 2015 that was published in the Journal of Retailing and Consumer Services and cited 244 times. The other four most cited studies of the sample were not anyhow connected to the mentioned group.

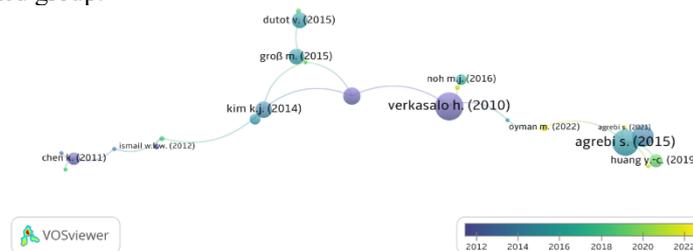

**Fig. 6.** Citation of TAM–related scientific works: visualization of the biggest group

In the case of UTAUT/UTAUT2, the largest set of documents consists of nine scientific works that might be combined into four categories; it is about 13% of the overall sample (Figure 7). The authors of these studies are illustrated in the figure below. Interestingly, this group included only the study of Okumus, Ali, Bilgihan, and Ozturk from 2018 which was published in the International Journal of Hospitality Management and cited 166 times. The other four most cited studies that applied UTAUT/UTAUT2 were not anyhow connected to the mentioned group.

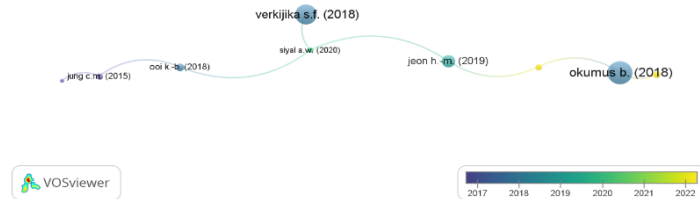

**Fig. 7.** Citation of UTAUT/UTAUT2–related scientific works: visualization of the biggest group

In the case of IDT, the largest set of documents consists of three scientific works that might be combined into four categories; it is about 20% of the overall sample (Figure 8). The authors of these studies are illustrated in the figure below. Interestingly, this group included two studies from Figure 8 that were conducted by Meng et al. [22] and Tsai and Ho [23]. The studies were cited 25 and 19 times respectively. The other three most cited works were not anyhow connected to the mentioned group.

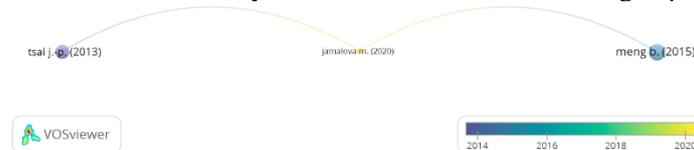

**Fig. 8.** Citation of IDT–related scientific works: visualization of the biggest group

### 5.3 Bibliographic coupling - TAM; UTAUT/UTAUT2; DOI

The answer to the third question is focused on applying VOSviewer to understand the number of shared references in two papers using bibliographic coupling [20]. It is also well-known that the mentioned method is one of the best tools for illustrating precise research areas [24]. To determine the most connected journals that might have a bigger impact on the development of the body of knowledge, the author set the minimum number of documents as three and the minimum number of citations as two.

Seven journals out of 98 were selected. The figure illustrates that some journals began to pay attention to the mentioned field a bit later than others (Figure 9). For example, the International Journal of Technology and Human Interaction, the Journal of Theoretical and Applied Information Technology as well as Cogent Business and Management mostly focused on TAM-related papers after 2019 while the International Journal of Interactive Mobile Technologies has been publishing scientific papers in the mentioned field since 2018.

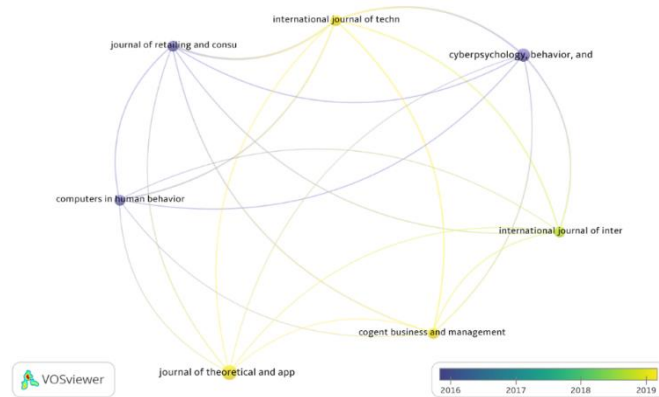

**Fig. 9.** Bibliographic coupling of TAM–related scientific works: visualization of relationships between 7 journals

In the case of UTAUT/UTAUT2-related studies, 5 journals out of 65 were selected. Figure 10 illustrates that some journals began to pay attention to the mentioned field a bit later than others. For example, the International Journal of Innovation and Technology Management as well as the Journal of Theoretical and Applied Information Technology were mostly focused on UTAUT/UTAUT2-related papers since 2020.

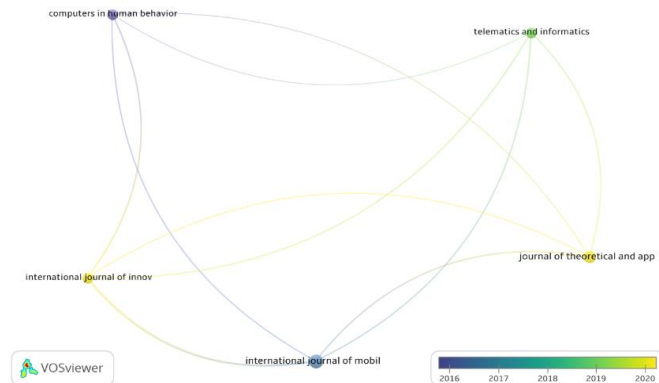

**Fig. 10.** Bibliographic coupling of UTAUT/UTAUT2 – related scientific works: visualization of relationships between 5 journals

In the case of IDT-related studies, 14 journals out of 19 were selected. Considering that only 19 papers were included in the analysis, the author set the minimum number of documents and the minimum number of citations as one. Figure 11 illustrates that some journals began to pay attention to the mentioned field a bit later than others. For example, the Computers in Human Behavior, Asia Pacific Journal of Tourism,

Industrial Management & Data Systems were mostly focused on IDT-related papers since 2014.

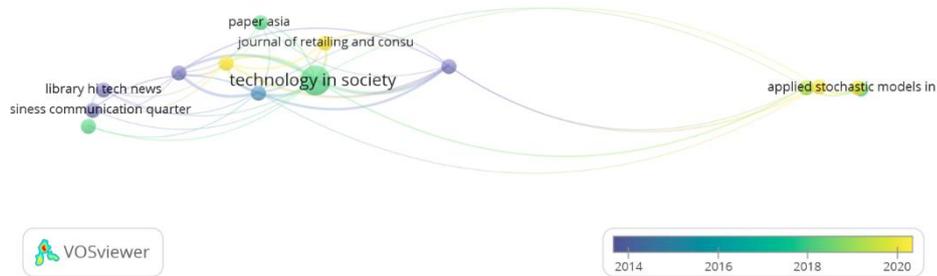

**Fig. 11.** Bibliographic coupling of IDT–related scientific works: visualization of relationships between 14 journals

### 5.4 Co-occurrence / co-word analyses - TAM; UTAUT/UTAUT2; DOI

To determine the development direction of the studies and the state of arts in the body of knowledge in general, the author conducted Co-occurrence analyses that were mentioned in the fourth research question. For this purpose, the unit of analysis was selected as all keywords, and the minimum number of keyword appearances was set as five. As a result, 24 keywords out of 659 were included in the analyses (Figure 12). Five factors were determined on the basis of the keywords. The most interesting keywords are social influence, social and economic effects, and trust.

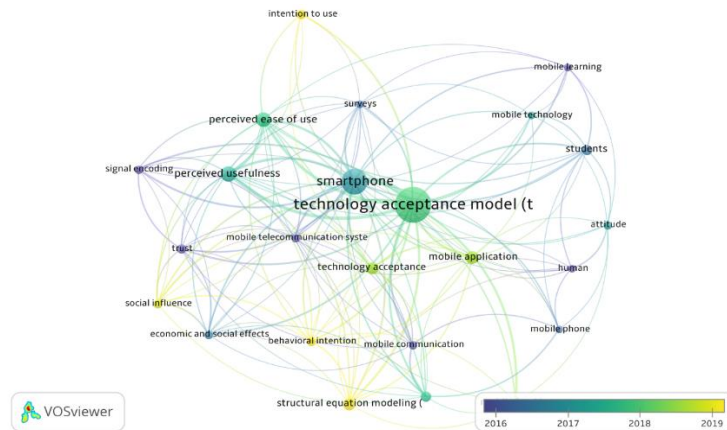

**Fig. 12.** Co-occurrence / co-word analyses of TAM–related scientific works: visualization of relationships between 24 keywords

The inclusion of the other keywords seems to be rather logical, and directly linked with TAM as well as the smartphone industry. Figure 11 illustrates that trust has been used in TAM-related models for quite a long time, nevertheless, the impact of social influence captured the attention of scientists since 2019. Also, after 2019 researchers mostly tend to use different versions of SEM in TAM-related calculations.

In the case of UTAUT / UTAUT2, 16 keywords out of 566 were included in the analyses (see Figure 13). Three factors were determined based on the keywords. The cooccurrence of keywords illustrates that research in developing countries is one of the distinguishing elements of the application of UTAUT / UTAUT2. The involvement of trust in the models as well as utilizing the model in the case of mobile learning is the similarities of UTAUT and TAM-affiliated research.

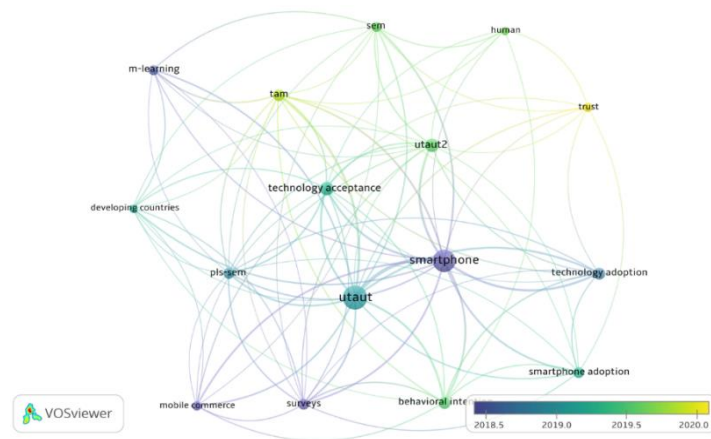

**Fig. 13.** Co-occurrence / co-word analyses of UTAUT/UTAUT2–related scientific works: visualization of relationships between 16 keywords

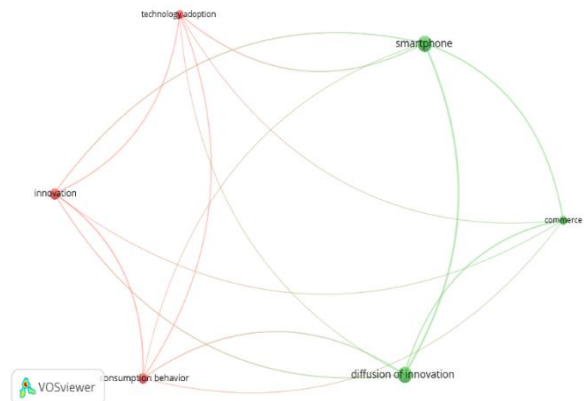

**Fig. 14.** Co-occurrence / co-word analyses of IDT–related scientific works: visualization of relationships between 6 keywords

In the case of IDT, 6 keywords out of 207 were included in the analyses (see Figure 14). Considering that only 19 papers were included in the analysis, the minimum number of keyword appearances was decreased to three. Two factors were determined on the basis of the keywords. The cooccurrence of keywords illustrates that research in commerce is one of the distinguishing elements of the application of DOI.

## 6  Discussion - Stage Two: Brief review of selected studies

### 6.1  Technology Acceptance Model for Smartphone Context

A brief history of TAM's development was explained above. In this section, the author would like to draw attention to the adoption of TAM in the smartphone adoption/diffusion context. The model is a simple yet powerful tool for explaining end user's behavior [25]. However, it was created to measure the behavior of individuals in the workplace [12], [26]. It means that essential elements focusing on the selection of devices (price-value, habit) and external factors (advertisements, social influence) that might have a big weight on customer decisions were not included in the model. Also, the original model does not consider any moderation effect that proved to be important in the case of technology diffusion [27]–[29].

One of the most influential papers that were mostly cited was published by Hsieh et al. [30] and focused on the understanding of learning English as a foreign language via LINE smartphone app. The authors applied a mixed-method research technique; quantitative data analyses were conducted using TAM among Taiwanese and exchange students. All the other papers included in Table 2 applied only quantitative research methods. The studies were published in the Journal of Computer Assisted Language Learning, Journal of Retailing and Consumer Services as well as two of Elsevier's journals (i.e., Telematics and Informatics Computers in Human Behavior). The highest number of citations was 281, while the other papers were cited 278 and 244 times respectively. The most cited studies using TAM were mainly focused on smartphone applications [30], [31], mobile shopping [32], mobile wallets [33], and smartphone dependency [34]. Mostly, all illustrated papers extended TAM by involving some variables such as technical barriers [31], perceived enjoyment and satisfaction [32], Mobile wallet self-efficacy, Informal learning, and Trust [33] as well as some other variables. However, co-occurrence analysis illustrated that researchers mainly apply TAM for understanding economic and social effects (including social influence), as well as the relationship between TAM variables in mobile learning/technology contexts.

### 6.2  Unified Theory of Acceptance and Use of Technology for Smartphone Context

In this section, the author would like to draw attention to the adoption of UTAUT/UTAUT2 in the smartphone adoption/diffusion context. It is clear that the model is a complex and powerful tool for explaining end user's behavior [35]. Even though the original version was created for measuring the behavior of individuals in the workplace [12], the extension is focused on the customer/end-user electronics context.

**Table 2.** A brief review of the most influential papers included in the analyses.

| Model : | Journal | Institution | Cited |
|---|---|---|---|
| **TAM** | | | |
| Chen Hsieh J.S., Wu W.-C.V., Marek M.W. - Using the flipped classroom to enhance EFL learning | Computer Assisted Language Learning (2016) /UK | National Central University (Taoyuan City) Taiwan; Providence University, (Taichung City), Taiwan; Wayne State College (NE), US | 281 |
| Verkasalo H., López-Nicolás C., Molina-Castillo F.J., Bouwman H. - Analysis of users and non-users of smartphone applications | Telematics and Informatics *(2010)/UK | Helsinki University of Technology, Finland; University of Murcia, Spain; Delft University of Technology, Netherlands | 278 |
| Agrebi S., Jallais J. - Explain the intention to use smartphones for mobile shopping | Journal of Retailing and Consumer Services (2015) / UK | France Business School (Tours Cedex), France; University of Rennes (Rue Jean Macé), France | 244 |
| Shaw N. - The mediating influence of trust in the adoption of the mobile wallet | Journal of Retailing and Consumer Services (2014) / UK | Ryerson University (Toronto), Canada | 154 |
| Park N., Kim Y.-C., Shon H.Y., Shim H. - Factors influencing smartphone use and dependency in South Korea | Computers in Human Behavior (2013)/UK | Yonsei University (Seoul), South Korea (SK); Cheil Worldwide, Seoul, South Korea; Korea Information Society Development Institute, | 140 |
| **UTAUT/UTAUT2** | | | |
| Tam C., Santos D., Oliveira T. - Exploring the influential factors of continuance intention to use mobile Apps: Extending the expectation confirmation model | Information Systems Frontiers (2020)/ NL | NOVA Information Management School (Lisbon), Portugal | 202 |
| Shaw N., Sergueeva K. - The non-monetary benefits of mobile commerce: Extending UTAUT2 with perceived value | Int. Journal of Information Management (2019)/UK | Ryerson University (Toronto), Canada | 201 |
| Okumus B., Ali F., Bilgihan A., Ozturk A.B. - Psychological factors influencing customers' acceptance of smartphone diet apps when ordering food at restaurants | International Journal of Hospitality Management (2018)/UK | University of Central Florida (FL), United States; University of South Florida Sarasota-Manatee (FL), United States; Florida Atlantic University, (FL), United States; | 166 |

| Qasim A., Abu-Shanab E. - Drivers of mobile payment acceptance: The impact of network externalities | Information Systems Frontiers (2015)/NL | Yarmouk University (Irbid), Jordan | 136 |
|---|---|---|---|
| Gupta A., Dogra N., George B. - What determines tourist adoption of smartphone apps?: An analysis based on the UTAUT-2 framework | Journal of Hospitality and Tourism Technology (2018)/UK | University of Jammu (Jammu), India; Fort Hays State University (Hays KS), United States | 128 |
| **IDT** | | | |
| Pham T.T.T., Ho J.C. - The effects of product-related, personal-related factors and attractiveness of alternatives on consumer adoption of NFC-based mobile payments | Technology in Society (2015)/UK | Yuan Ze University (Chung-Li), Taiwan | 171 |
| Kaur P., Dhir A., Bodhi R., Singh T., Almotairi M. - Why do people use and recommend m-wallets? | Journal of Retailing and Consumer Services (2020)/UK | Aalto University, Finland; North-West University, (Vanderbijlpark), South Africa; Lappeenranta University of Technology (Lappeenranta), Finland; Motilal Nehru National Institute of Technology (Allahabad), India; King Saud University, SA | 70 |
| Meng B., Kim M.-H., Hwang Y.-H. - Users and Non-users of Smartphones for Travel: Differences in Factors Influencing the Adoption Decision | Asia Pacific Journal of Tourism Research (2015)/UK | Shanxi University (Taiyuan City), China; Dong-A University (Busan), South Korea | 25 |
| Zhang X. - Frugal innovation and the digital divide: Developing an extended model of the diffusion of innovations | International Journal of Innovation Studies (2018)/UK | University of North Texas (TX), United States | 24 |
| Tsai J.-P., Ho C.-F. - Does design matter? Affordance perspective on smartphone usage | Industrial Management & Data Systems (2013)/UK | Far East University (Tainan), Taiwan; National Sun Yat-Sen University (Kaohsiung), Taiwan | 19 |

Source: own editing based on literature review.

It means that essential elements focusing on smartphone selection [36] as well as external factors (social influence) are included in the model. Also, the application of moderators might increase the exploratory power of the model [15], [29]. Almost all the studies excluding [37] listed in Table 2 used the extended version of UTAUT and involved some new variables. These variables are perceived value [38], confirmation and satisfaction [39], personal innovativeness [40], trust and network externality [37], perceived risk, and perceived trust [41]. So, the mentioned authors attempt to extend the current model and increase the body of knowledge with new findings. Two of the mentioned papers were published in Information Systems Frontiers [37], [39]. The papers mostly focus on understanding intention towards mobile apps [39] including diet apps [40] and tourist apps [41] as well as mobile payment [37] and commerce [38]. However, co-occurrence analysis proved that researchers mainly apply UTAUT/UTAUT2 for understanding mobile learning, and mobile commerce with a strong focus on developing country context.

### 6.3 Innovation Diffusion Theory for Smartphone Context

In this section, the author would like to draw attention to the adoption of IDT in the smartphone adoption/diffusion context. The model is a complex and powerful tool for explaining end user's behavior [17]. Moreover, Rogers [42] paid special attention to the field of telecommunication where four main elements of diffusion (innovation, channels, time, and members) impact the above-mentioned model. Also, the theory summarizes the factors influencing the acceptance of innovative products under personal, social, and technological categories [43-44]. However, as the models highlighted before, IDT is also continuously improved and extended [45]. Based on the results of the literature review (see Table 2), IDT was mainly applied in the context of mobile payments (i.e., m-wallet [21] and NFC payments [46]) as well as smartphone adoption-related topics (travel [1]; smartphone [23], [47], agriculture [48]). Co-occurrence analysis [49] also illustrated that researchers usually apply IDT to understand payments and commerce.

## 7 CONCLUSION

The result of bibliometric analyses as well as the literature review shows that all three models might apply to the mentioned topic. However, the unit of analysis might be an essential point to determine which of the models fits the scope of analyses better. Moreover, the most cited TAM-related papers were published between 2010 and 2017; in the case of UTAUT / UTAUT2 significant scientific works were released from 2016 to 2020. Citation linkage results illustrate that most of TAM-related papers were linked to each other. It might be connected to the fact that TAM is well known tool in comparison to UTAUT/UTAUT2 which is a new model. The visualization of data also illustrates that IDT is the least utilized model.

Co-occurrence / co-word analyses of TAM proved studies mostly conducted among students and the model was extended by adding variables focusing on social influence, social/economic effects, and trust. The cooccurrence of keywords in the case of UTAUT/UTAUT2 illustrates that the research in developing countries is one of the distinguishing elements. The involvement of trust in the models as well as utilizing the model in the case of mobile learning is the similarities of UTAUT and TAM-affiliated research. Nevertheless, in the case of IDT, the diffusion of innovations was the main issue.

Co-occurrence analyses proved that studies involved in the analyses almost did not use moderating and mediation effects which is one of the biggest issues. Also defining the moderation effect of age and gender as well as the mediation effect of other variables will enrich the literature and decrease the knowledge gap. In the future, it would be better to check not only separate relationships but also try to determine the general model for understanding behavior towards the adoption of smartphones/wearable devices in the wider context.